\label{key}
\documentclass[graphicx,amsmath,amssymb,prl,twocolumn]{revtex4-1}

\usepackage{graphicx}		
\usepackage{color}
\usepackage{amsmath}
\usepackage{subfigure}
\usepackage{upgreek}

\usepackage{dsfont}

\begin{document}
	
	\title{The sound of an axon's growth}
	
	\author{Frederic Folz}
	\affiliation{Theoretische Physik, Universit\"at des Saarlandes, 66041 Saarbr\"ucken, Germany}
	\author{Lukas Wettmann}
	\affiliation{Theoretische Physik, Universit\"at des Saarlandes, 66041 Saarbr\"ucken, Germany}
	\author{Giovanna Morigi}
	\affiliation{Theoretische Physik, Universit\"at des Saarlandes, 66041 Saarbr\"ucken, Germany}
	\author{Karsten Kruse}
	\affiliation{NCCR Chemical Biology, Departments of Biochemistry and Theoretical Physics, University of Geneva, 1211 Geneva, Switzerland}
	
	\date{\today}
	
	\begin{abstract} 
		Axons are linear structures of nerve cells that can range from a few tens of micrometers up to meters in length. In addition to external cues, the length of an axon is also regulated by unknown internal mechanisms. Molecular motors have been suggested to generate oscillations with an axon-length dependent frequency that could be used to measure an axon's extension. Here, we present a mechanism for determining the axon length that couples the mechanical properties of an axon to the spectral decomposition of the oscillatory signal. 
	\end{abstract}
	
	\maketitle
	
	In order to assure proper function, the size of a biological system typically needs to be regulated. There is currently no general understanding of the underlying mechanisms. Best studied are processes based on gradients for setting the extension of linear structures. Prominent examples are provided by the length of cytoskeletal filaments~\cite{Varga:2006er,Varga:2009jy,Johann:2012kk,Melbinger:2012hp,Erlenkamper:2013cy,Mohapatra:2015cc,Mohapatra:2016kq,Fritzsche:2016fx} and the extension of antiparallel microtubule overlaps~\cite{Channels:2008io,Johann:2015da,Braun:2017hx}. In addition, length-dependent mechanical forces can play a role as has been suggested for setting the size of stereocilia~\cite{Prost2007}. Finally, oscillations are involved in positioning the division plane of some bacteria~\cite{Loose:2011dd,Wettmann:2018bp} and thereby determine the size of the daughter cells. They may be viewed as a specific case of cavity resonances of chemical waves that have been proposed to provide a general mechanism for size determination of biological structures~\cite{Laughlin:2015hm}. 
	
	Axons are linear structures along which electrical signals emanating from the body (soma) of a nerve cell are transported to other cells. The length of an axon can vary from a few micrometers up to meters. It is set in part by extrinsic mechanisms, for example, stretch growth: axons that have connected to other cells are experiencing mechanical tension as the organism is growing, which induces axonal extension~\cite{Athamneh:2015ba}. Prior to making contact with other cells and driven by a structure called the growth cone, axons extend at their tip. Growth cones are guided in part by external physical~\cite{Koser:2016ks} and chemical cues~\cite{Erskine:2007dw}. In addition, there are intrinsic mechanisms to set the axon length that are notably used in early stages of organismal development~\cite{Albus:2013es}. 
	
	\begin{figure}
		\includegraphics[width=0.45\textwidth]{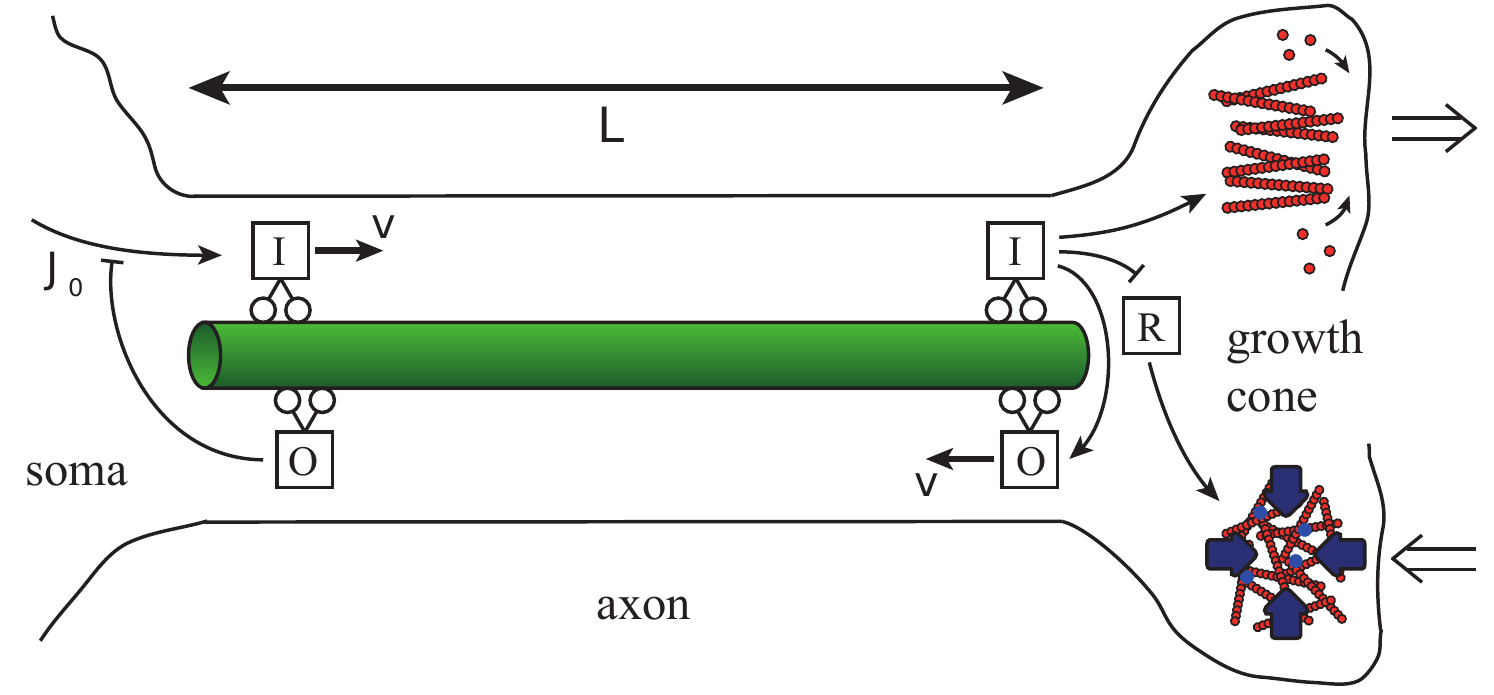}
		\caption{\label{fig:schema}(color online) Illustration of axon length regulation. Motors transport signals $I$ and $O$ along microtubules at velocity $v$, respectively, from the soma to the growth cone and \textit{vice versa}. Activation of $O$-transport in the growth cone by $I$ and suppression of $I$ at the soma by $O$ generate oscillations with a frequency depending on the axon length $L$. The total influx of $I$ comprises both a constant $J_0$ corresponding to the maximal influx of motors and the inhibitory effect of $O$. Furthermore, the signal $I$ also stimulates actin polymerization leading to growth cone extension, as well as inhibition of the response $R$. In turn, $R$ induces actin network contraction leading to growth cone retraction. The extending and contracting actin networks are drawn separately for clarity, but can be colocalized. Lines with arrow heads indicate activation/stimulation, lines with blunt ends inhibition.}
	\end{figure}
	It has been shown that transport by molecular motors -- with kinesins moving from the soma to the axon tip and cytoplasmic dynein in the opposite direction -- is essential for intrinsic axonal length regulation~\cite{Myers:2007ig,Rishal:2012jg,Perry:2016he}. However, the precise role of the motors in this process is currently unknown. Similar to motor-dependent length-regulation of filaments~\cite{Varga:2006er,Varga:2009jy,Johann:2012kk,Melbinger:2012hp,Mohapatra:2015cc}, one possibility is that they generate a gradient along the axon. Yet, a gradient-based mechanism is unlikely to operate over more than a few micrometers in cells, whereas in developing embryos, axons up to a few hundred micrometers can emerge~\cite{Albus:2013es}. Furthermore, decreased motor concentrations would imply shortened axons. However, the opposite is observed~\cite{Myers:2007ig,Rishal:2012jg}. These considerations have led to the proposal that interactions between motors moving in opposite directions along the axon generate an oscillating signal with a length-dependent frequency~\cite{Rishal:2012jg}. Indeed, if kinesins transported some signaling molecule to the tip, where it initiated dynein-mediated transport of another signalling molecule that in turn stopped the original chemical signal at the soma, then the concentrations of these factors at the growth cone and at the soma can oscillate~\cite{Karamched:2015ha}, see Fig.~\ref{fig:schema}. For motors with a constant velocity the associated period would increase proportional to the axon length. However, it is not clear how this frequency-dependence could be used for length regulation. One proposition is that there is a network generating a frequency-dependent average of some signaling molecule coupled to a switch turning off further axonal growth as a certain concentration threshold and therefore length is reached~\cite{Bressloff:2015gg}. 
	
The actual growth dynamics of an axon regulated by length-dependent oscillations has not yet been studied. Here, we propose a mechanism for this regulation that couples the oscillations to the axon's mechanical properties. Axons are known to be under mechanical tension: on one hand, the growth cone pulls on the axon~\cite{Bray:1979tg}, on the other hand, the axon itself generates contractile mechanical stress by the actin cytoskeleton~\cite{Ahmad:2000wv,Bernal:2007gx}. The regulation of cytoskeletal stresses has been suggested to generate bouts of elongation and retraction~\cite{Baas:2001kk}. The oscillations can be analyzed through differential equations with delayed feedback~\cite{Karamched:2015ha}.
		
		We start from these equations and explicitly include the axon growth dynamics, obtaining a set of equations with a state-dependent delay. We show that our mechanism, rather than reading out the oscillation frequency, exploits the information contained in the signal's spectral composition for regulating axon growth and length. Notably, we find regions in parameter space where a reduction of the motor concentration leads to an increase in the final axon length, consistent with experimental findings~\cite{Myers:2007ig,Rishal:2012jg}.
	
	Let us begin our discussion by noting that, typically, the cellular response to a chemical signal shows a sigmoidal dose-response curve~\cite{Alon06:nn}. It is often given in terms of a Hill function
	\begin{align}
	f_\kappa\left(c\right)&=\frac{c^n}{\kappa^n+c^n}
	\end{align}   
	with Hill coefficient $n$ and half-concentration $\kappa$. For an oscillating chemical signal, the average response is independent of the signal's period $T$: Let $c_I$ denote the concentration of an incoming signaling molecule with $c_I(t+T)=c_I(t)$. The response $c_R$ is then given by 
	\begin{align}
	\label{eq:cRdot}
	\dot{c}_R &= J_R(1-f_\kappa\left(c_I\right)) - \gamma_R c_R ,
	\end{align}
	where $J_R$ is the coupling constant between $c_I$ and $c_R$ and $\gamma_R$ is the response's decay rate. 
	In the following we will scale the concentrations by $\kappa$ and denote them by the same symbols as before. The average response is then $\langle c_R\rangle \equiv \frac{1}{T}\int_0^T c_R(t)\;\mathrm{d}t=\bar{J}_R(1-\langle f_1\left(c_I\right)\rangle)$, which is scaled by the dimensionless parameter $\bar{J}_R\equiv J_R/(\gamma_R\kappa)$. It holds that $\frac{1}{T}\int_0^{T} f_1\left(c_I( t)\right) \;\mathrm{d}t=\frac{\beta}{T}\int_0^{T/\beta} f_1\left(c_I(\beta t)\right) \;\mathrm{d}t$ for any $\beta>0$ showing that $\langle c_R\rangle$ is independent of the oscillation period. 
	
	This general result can be illustrated in the limit $n\to\infty$, when the Hill function turns into a Heaviside function, and one can determine the time dependence of $c_R$ explicitly. Let the input signal be some oscillatory function with exactly one minimum and one maximum per period $T$ and let $t=0$ and $t=t_\times$ be the times, when the input signal equals the threshold, such that $c_I(t)>1$ for $0\le t< t_\times$ and $c_I(t)<1$ for $t_\times\le t<T$. Then the average is given by 
	\begin{align}
	\label{eq:cR:ave}
	\langle c_R\rangle &=\bar{J}_R\left(1-\frac{t_\times}{T}\right)\,.
	\end{align}
	Apart from the coupling constant $\bar{J}_R$, $\langle c_R\rangle$ only depends on the fraction of the period during which the incoming signal $c_I$ is larger than $1$, but is independent of the period. 
	
	Whereas $\langle c_R\rangle$ does not depend on the frequency of $c_I$, the form, i.e., the spectrum of $c_R$ does. In turn, $\langle c_R\rangle$ typically depends on the spectrum of $c_I$. This is easily seen if $f_\kappa$ is a Heaviside function as in the previous paragraph: As long as $\min c_I<\kappa<\max c_I$, a change in the spectrum of $c_I$ typically changes the fraction of time $c_I>\kappa$ and hence $\langle c_R\rangle$. Consequently, by first transforming the frequency variation of a signal $O$ into a variation of the shape of the incoming signal $I$ via a process similar to Eq.~(\ref{eq:cRdot}) and then reading out its shape via Eq.~(\ref{eq:cRdot}), variations in frequency can be ultimately turned into a variation of the average value of the response $R$.
	\begin{figure}
		\includegraphics[width=0.47\textwidth]{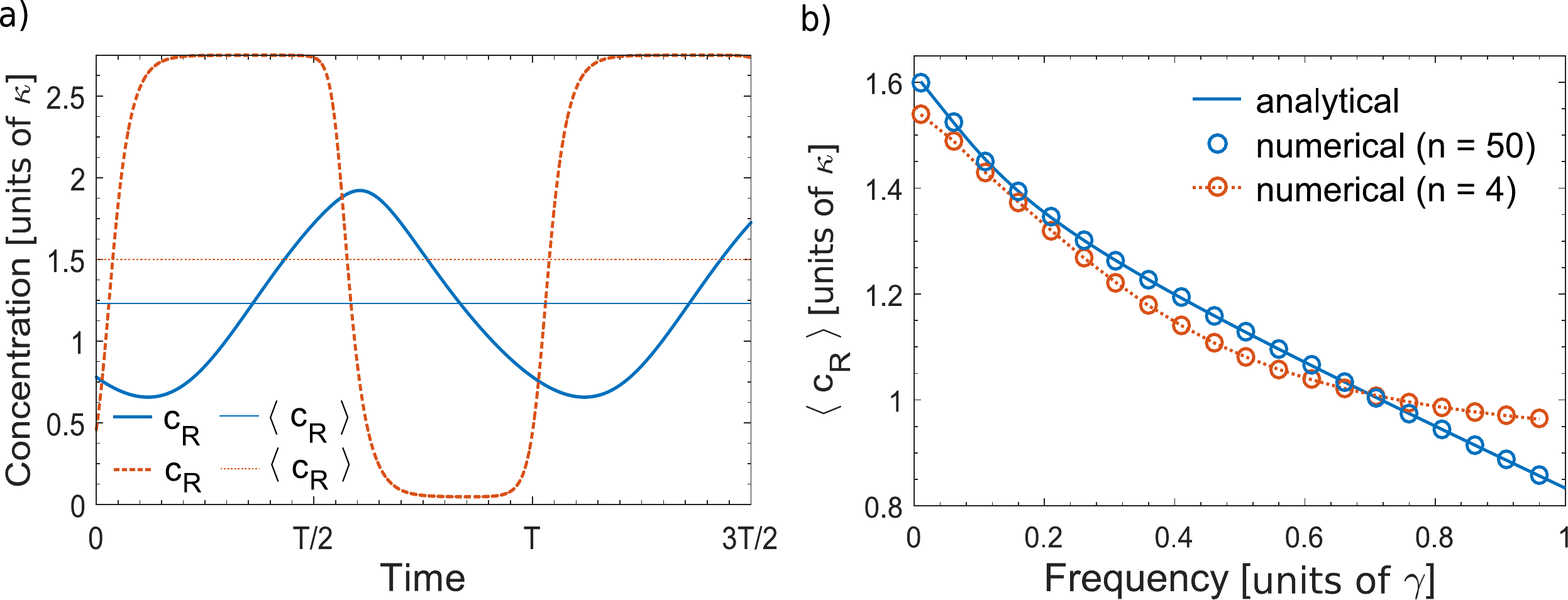}
		\caption{\label{fig:averageAsFunctionOfFrequency}(color online) Two-step transformation of a periodic signal into a frequency-dependent average response, Eqs.~(\ref{eq:cRdot}) and (\ref{eq:cGdot}). a) Dynamics of the response $c_R$ as a function of time and corresponding mean value $\langle c_R\rangle$ for two different frequencies of a sinusoidal oscillatory signal $O$, $c_{O}(t) = \bar J_R(1+\sin(2 \pi t/T))/2$. b) Response average value $\langle c_R\rangle$ as a function of the oscillatory signal frequency for two different Hill coefficients. Full line represents the analytic expression Eq.~(\ref{eq:cR:ave}) with $t_\times$ replaced by $t_>$. Parameter values are $J_R = J_I = 55 \times 10^{-5} \,\upmu$m$^{-1}$s$^{-1}$, $\kappa = 2 \times 10^{-2} \,\upmu$m$^{-1}$, $\gamma = 10^{-2}\,$s$^{-1}$ as well as $n=4$ (a) and $n=4$ (dotted line with circles) and $n=50$ (full line with circles) in panel (b).}
	\end{figure} 
	
	To give a specific illustration of this mechanism, consider an oscillatory signal $O$, which feeds into $I$ through
	\begin{align}
	\label{eq:cGdot}
	\dot c_I &= J_I \left(1-f_\kappa(c_O)\right)-\gamma_Ic_I,
	\end{align} 
	where $J_I$ is the coupling constant between $O$ and $I$ and $\gamma_I$ is the decay rate of the incoming signal. Furthermore, we used the same sigmoidal function $f_\kappa$ in Eqs.~(\ref{eq:cRdot}) and (\ref{eq:cGdot}), but our general results do not rely on these specific choices. We will also choose $J_I=J_R$ and $\gamma_I=\gamma_R\equiv\gamma$ to not blur the general mechanisms by a multitude of parameters. As above we will scale the densities by $\kappa$. As anticipated, the incoming signal $I$ responds to frequency changes in $O$ by variations of its shape, see Fig.~\ref{fig:averageAsFunctionOfFrequency}a. Furthermore, the average value of the eventual response $R$ varies with the frequency of $O$, see Fig.~\ref{fig:averageAsFunctionOfFrequency}b. In case $f_\kappa$ is a Heaviside function, the average of $c_R$ as a function of the period $T$ of $c_O$ takes the same form as Eq. \eqref{eq:cR:ave} with the replacement $t_\times\to t_{>}$. Here, $t_{>}$ is the length of the time interval during which $c_I>1$, where $t_{>}>0$ requires $\bar{J}_I>1$. The value of $t_>$ is determined by a transcendent equation, which we omit here. The average response increases with $T$ and $\langle c_R\rangle\to \bar{J}_R (1 - t_>/T)$ for $n\to\infty$. 
	
	Having established a mechanism for reading out the frequency of the incoming signal, we now return to axon length regulation. As mentioned in the introduction, motors moving in opposite directions and carrying motor-transport activating or inhibiting signals can generate axon-length dependent oscillations. Specifically, let $I$ denote a signal released from the soma and transported into the growth cone. Its concentration at the growth cone is denoted by $c_I(t)$. In addition to eliciting the response $R$, it also triggers transport of an outgoing signal $O$ from the growth cone to the soma. There, $O$ suppresses further transport of $I$. The concentration of $O$ at the soma is denoted by $c_O(t)$. This dynamics can be captured as in Ref.~\cite{Karamched:2015ha} by the following delay-differential equations
	\begin{align}
	\label{eq:cIdot}
	\dot c_I(t) & = J_0 - J_I f_\kappa\left(c_O(t-\tau)\right) - \gamma_I c_I(t) \\
	\label{eq:cOdot}
	\dot c_O(t) & =  J_O f_\kappa\left(c_I(t-\tau)\right)-\gamma_O c_O(t) .
	\end{align}
	The delay $\tau=L/v$ accounts for the time motors need to transport cargo along an axon of length $L$ at a velocity $v$, which we assume for simplicity to be the same for both kinds of motors. Note that for $\tau=0$, Eq.~(\ref{eq:cIdot}) is the same as Eq.~(\ref{eq:cGdot}) if $J_0=J_I$. The incoming signal decays at rate $\gamma_I$ at the growth cone and the outgoing signal at rate $\gamma_O$  at the soma. $J_0$ is the maximal incoming flux of signal $I$ and $J_I\le J_0$ as well as $J_O$ denote the respective coupling constants between $O$ and $I$. To avoid unnecessary complications, we consider again the same sigmoidal function $f_\kappa$ as in Eqs.~(\ref{eq:cRdot}) and (\ref{eq:cGdot}) and focus on the case $\gamma_I=\gamma_O\equiv\gamma$ and $J_0=J_I=J_O\equiv J$. 
	
\begin{figure}
		\includegraphics[width=0.5\textwidth]{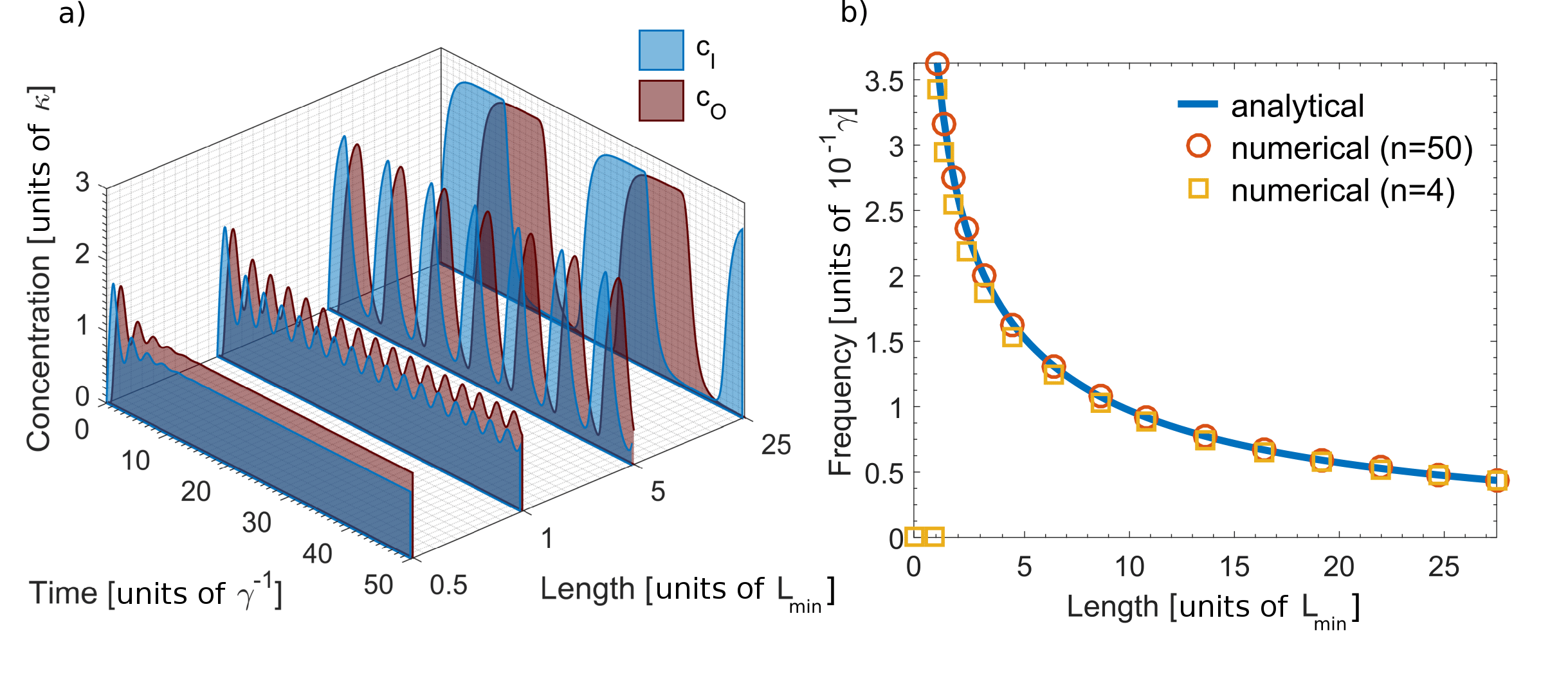}
		\caption{\label{fig:motorOscillations}(color online) Motor-induced oscillations for fixed axon length. a) Oscillation of the incoming and outgoing signals $I$ and $O$, respectively, obtained by solving Eqs.~(\ref{eq:cIdot}) and (\ref{eq:cOdot}). b) Frequency of the oscillations generated by Eqs.~(\ref{eq:cIdot}) and (\ref{eq:cOdot}). The full line is obtained from Eq.~(\ref{eq:omegaAsFunctionOfL}). Parameter values as in Fig.~\ref{fig:averageAsFunctionOfFrequency} and $J_0 = J_O = J_I$, $\gamma_O = \gamma_I = \gamma$ as well as $n = 4$ (a) and  $n=4$ (squares) and $n=50$ (circles) in panel (b).}
	\end{figure}

	For fixed length $L$, a linear stability analysis of the stationary state $c_{I,0}$ and $c_{O,0}$ of Eqs.~(\ref{eq:cIdot}) and (\ref{eq:cOdot}) shows that the system undergoes a Hopf-bifurcation when the parameters fulfil
	\begin{align}
	\label{Lmin}
	\bar{J}\sin\left(\gamma\tau\sqrt{\bar{J}^2-1}\right)&=1,
	\end{align}
	where $\bar J=\sqrt{\alpha} J/(\kappa\gamma)$ and $\alpha=\langle f_1'(c_O)\rangle\langle f_1'(c_I)\rangle$~\cite{Karamched:2015ha}. In particular,  $f_1'(x)=df_1(x)/dx$ and the mean values are taken at the instability point $c_I=c_I^*$, $c_O=c_O^*$, cf. Supplementary Material (S.M.). This equation can only be fulfilled if $\bar{J}>1$. The linear stability analysis also shows that there is a minimal axon length $L_\mathrm{min}=v\tau_\mathrm{min}$, below which the system does not oscillate, which is in agreement with numerical solutions of Eqs.~(\ref{eq:cIdot}) and (\ref{eq:cOdot}), see Fig.~\ref{fig:motorOscillations}a. The frequency $\omega_\mathrm{min}$ at this critical axon length is finite and fulfils 
	\begin{align}
	\cot\left(\omega_\mathrm{min}\tau_\mathrm{min}\right)&=\frac{\omega_\mathrm{min}}{\gamma}\,.
	\end{align}
	Remarkably, this relation between the axon length and the oscillation frequency also determines the frequency of the full nonlinear oscillations, see Fig.~\ref{fig:motorOscillations}b. As a function of the axon length $L$ it is approximately given by 
	\begin{align}
	\label{eq:omegaAsFunctionOfL}
	\omega\sim\frac{1}{\sqrt{\frac{L}{v}\left(\frac{L}{3v}+\gamma^{-1}\right)}}.
	\end{align}
	For $L\gamma\gg v$, we have $\omega \approx \sqrt{3}v/L$, which is the solution of a wave equation with a rescaled sound velocity. In the opposite limit, the frequency scales as $\omega\approx \sqrt{v\gamma/L}$.  
	
	Whereas the previous analysis was for a fixed axon length, we will now consider $L$ to be a dynamic variable. The axon length is regulated by two processes: an extension of the growth cone and a shortening of the axon due to contractile stresses~\cite{Ahmad:2000wv,Bernal:2007gx}. The growth cone moves forward by a process similar to mesenchymal cell migration on a flat substrate: Extension of the leading edge is driven by the polymerization of actin, which is anchored to a large actin network and thus able to exert protruding forces on the membrane. The protrusion velocity is regulated by various processes. We assume here that chemical regulation through signal $I$ dominates and write for the protrusion velocity $v_gc_I$. We notably neglect an effect of membrane tension on the protrusion velocity~\cite{Mueller:2017br}. 
	
	The contractile stresses that are generated by molecular motors in the axon and the growth cone can be captured phenomenologically by a term $\zeta\Delta\mu$~\cite{Kruse:2004il}. Here, $\Delta\mu\equiv\mu_\mathrm{ATP}-\mu_\mathrm{ADP}-\mu_\mathrm{P}$, where $\mu_\mathrm{ATP}$ is the chemical potential of Adenosine-triphosphate, $\mu_\mathrm{ADP}$ that of Adenosine-diphosphate, and $\mu_\mathrm{P}$ that of inorganic phosphate, such that $\Delta\mu$ is the chemical energy liberated during an event of ATP-hydrolysis. The phenomenological coefficient $\zeta$ describes the coupling of the liberated chemical energy to the mechanical stress generated. For contractile stresses $\zeta<0$. The phenomenological coefficient $\zeta$ depends on regulatory signals~\cite{Bois:2011kx}. In particular, we assume $\zeta\equiv\zeta(c_R)$. Contractile stresses tend to reduce the distance between the cell body and the growth cone. We assume the cell body to be anchored to the substrate, such that contractile stresses are balanced by dissipative forces as the growth cone retracts. The latter can be written as $\xi \dot{x}_\mathrm{gc}=\zeta\Delta\mu$, where $x_\mathrm{gc}$ denotes the position of the growth cone and which we identify with the axon length $L$. Assuming a linear dependence of $\zeta$ on $c_R$, $\zeta=\zeta_1c_R$, and adding the growth cone protrusion velocity to the velocity due to contraction, we arrive at
	\begin{align}
	\label{eq:Ldot}
	\dot{L}&=v_gc_I-v_sc_R,
	\end{align}
	where $v_s=-\zeta_1\Delta\mu/\xi>0$. The response $R$ still depends on $I$ through Eq.~(\ref{eq:cRdot}). We will scale the length by $L_\mathrm{min}$ and concentrations by $\kappa$, while keeping the same notation for the axon length as well as for the growth and shrinkage velocities. Furthermore, we take $\gamma_R=\gamma$.
	
\begin{figure}
		\includegraphics[width=0.47\textwidth]{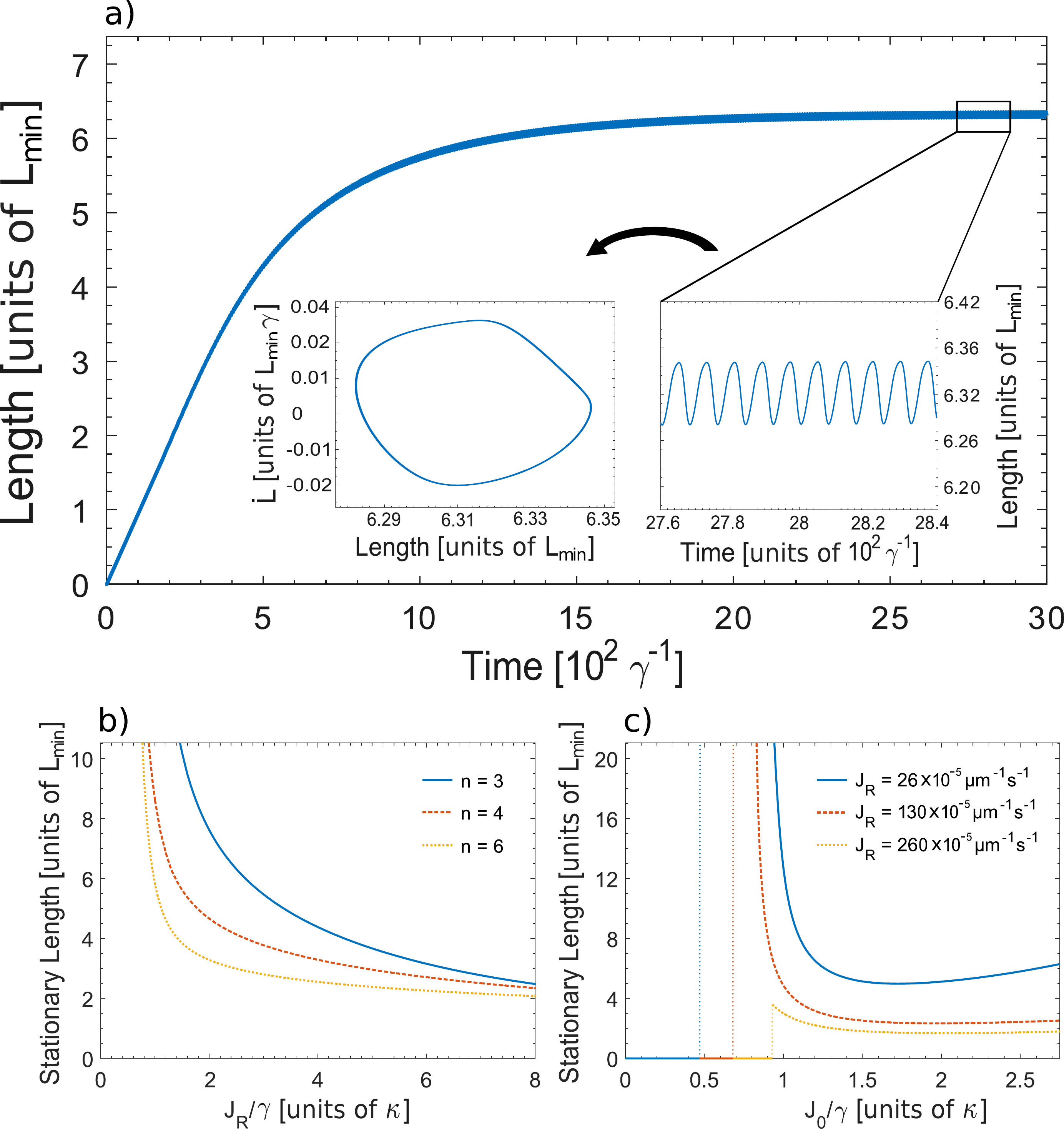}
		\caption{\label{fig:exampleLengthRegulation}(color online) 
			Axon length dynamics. a) Solution to the dynamic equations~(\ref{eq:cRdot}), (\ref{eq:cIdot}), (\ref{eq:cOdot}), and (\ref{eq:Ldot}). Parameter values are $J_R = 26 \times 10^{-5} \,\upmu$m$^{-1}$s$^{-1}$, $J = 55 \times 10^{-5}\,\upmu$m$^{-1}$s$^{-1}$, $\kappa = 2 \times 10^{-2} \,\upmu$m$^{-1}$, $\kappa_R = 5 \times 10^{-3} \,\upmu$m$^{-1}$, $\gamma = 10^{-2}\,$s$^{-1}$, $v_g = 0.1 \,\,\upmu$m$^2$s$^{-1}$, $v_s = 0.5 \,\upmu$m$^2$s$^{-1}$ and $n = 4$. b) Dependence of the mean length on $J_R$ for different values of the Hill coefficient $n$. c) Dependence of the mean length on $J_0$ for different values of the coupling constant $J_R$ and $J_I = J_0$. The other parameters are as in (a), $L_{\text{min}}$ is calculated from Eq.~\eqref{Lmin} with $J_I = 55 \times 10^{-5} \,\upmu$m$^{-1}$s$^{-1}$.}
	\end{figure}

	In Figure~\ref{fig:exampleLengthRegulation}a, we present an example of the solution to the dynamic equations (\ref{eq:cRdot}), (\ref{eq:cIdot}), (\ref{eq:cOdot}), and (\ref{eq:Ldot}). Starting from length zero, the length increases and eventually oscillates with an amplitude that is less than $1\%$ of the average length. The final average length decreases with an increasing coupling parameter $J_R$, Fig.~\ref{fig:exampleLengthRegulation}b.  This is because an increase of the coupling between the incoming signal and actomyosin contractility will increase the latter, which opposes the extension of the axon. A similar effect is observed when reducing the actomyosin activity, which is generally achieved by decreasing $\zeta_1\Delta\mu$ . Importantly, in Fig.~\ref{fig:exampleLengthRegulation}c it is visible that for sufficiently small values of $J_0$ the stationary length increases with decreasing $J_0$, that is with decreasing kinesin motor concentration.
This is  consistent with experimental findings~\cite{Myers:2007ig,Rishal:2012jg}. Only beyond a certain critical value of $J_0$ the average final length increases with an increasing motor concentration~\footnote{These results also hold if the value of $J_I$ is fixed while changing $J_0$.}. An increase of the axon lengths has been also observed when the dynein concentration is reduced~\cite{Rishal:2012jg}. Accordingly, our model shows an increase in the stationary length, when the parameter $J_O$ is reduced, see S.M. 
	
We have checked that the solutions are stable against fluctuations of the delay time up to 10\%, see S.M. We also found that lossy transport of the signals along the axon does not qualitatively change the system's behavior, as we show in the S.M.
	
	From a more general point of view, our system achieves length regulation through an adaptive delay, which is determined by the axon length. Adaptive delays have also been proposed as a mechanism to retrieve information from chaotic neural networks~\cite{Crook:2003ci}. It is also interesting to compare our mechanism to Laughlin's proposal of regulating lengths by resonant chemical waves~\cite{Laughlin:2015hm}, where he exploits a formal analogy of excitable systems with an effective amplifying and saturable medium, like in lasers or electronic circuits. Even though a mapping from the present model to reaction-diffusion equations is not evident, let us exploit the similarity of the term of length growth and shrinkage in Eq.~(\ref{eq:Ldot}) with gain and loss terms of a laser. As the system tunes its length, it eventually reaches a state in which the gains on average equal the losses. This goes hand in hand with the selection of a specific frequency and phase-locking of the different oscillating signals, which is akin to a resonance in an amplifying medium. 
	
	For the parameters chosen in Fig.~\ref{fig:exampleLengthRegulation}a, the average axon length would be 125~$\upmu$m and the oscillation amplitude below 1~$\upmu$m. Such small oscillations would likely be masked by fluctuations in a real axon. It would now be interesting to link in molecular detail the regulation of growth cone extension and axon contraction. In particular, the regulation of axon contractility most likely requires a description that explicitly accounts for the spatial degree of freedom along the axon. 
	
	\acknowledgments
	L.W. and K.K. acknowledge support by DFG through SFB 1027.


\begin{thebibliography}{99}
\expandafter\ifx\csname natexlab\endcsname\relax\def\natexlab#1{#1}\fi
\expandafter\ifx\csname bibnamefont\endcsname\relax
  \def\bibnamefont#1{#1}\fi
\expandafter\ifx\csname bibfnamefont\endcsname\relax
  \def\bibfnamefont#1{#1}\fi
\expandafter\ifx\csname citenamefont\endcsname\relax
  \def\citenamefont#1{#1}\fi
\expandafter\ifx\csname url\endcsname\relax
  \def\url#1{\texttt{#1}}\fi
\expandafter\ifx\csname urlprefix\endcsname\relax\def\urlprefix{URL }\fi
\providecommand{\bibinfo}[2]{#2}
\providecommand{\eprint}[2][]{\url{#2}}

\bibitem[{\citenamefont{Varga et~al.}(2006)\citenamefont{Varga, Helenius,
  Tanaka, Hyman, Tanaka, and Howard}}]{Varga:2006er}
\bibinfo{author}{\bibfnamefont{V.}~\bibnamefont{Varga}},
  \bibinfo{author}{\bibfnamefont{J.}~\bibnamefont{Helenius}},
  \bibinfo{author}{\bibfnamefont{K.}~\bibnamefont{Tanaka}},
  \bibinfo{author}{\bibfnamefont{A.~A.} \bibnamefont{Hyman}},
  \bibinfo{author}{\bibfnamefont{T.~U.} \bibnamefont{Tanaka}},
  \bibnamefont{and} \bibinfo{author}{\bibfnamefont{J.}~\bibnamefont{Howard}},
  \bibinfo{journal}{Nat.~Cell Biol.} \textbf{\bibinfo{volume}{8}},
  \bibinfo{pages}{957} (\bibinfo{year}{2006}).

\bibitem[{\citenamefont{Varga et~al.}(2009)\citenamefont{Varga, Leduc, Bormuth,
  Diez, and Howard}}]{Varga:2009jy}
\bibinfo{author}{\bibfnamefont{V.}~\bibnamefont{Varga}},
  \bibinfo{author}{\bibfnamefont{C.}~\bibnamefont{Leduc}},
  \bibinfo{author}{\bibfnamefont{V.}~\bibnamefont{Bormuth}},
  \bibinfo{author}{\bibfnamefont{S.}~\bibnamefont{Diez}}, \bibnamefont{and}
  \bibinfo{author}{\bibfnamefont{J.}~\bibnamefont{Howard}},
  \bibinfo{journal}{Cell} \textbf{\bibinfo{volume}{138}}, \bibinfo{pages}{1174}
  (\bibinfo{year}{2009}).

\bibitem[{\citenamefont{Johann et~al.}(2012)\citenamefont{Johann, Erlenkaemper,
  and Kruse}}]{Johann:2012kk}
\bibinfo{author}{\bibfnamefont{D.}~\bibnamefont{Johann}},
  \bibinfo{author}{\bibfnamefont{C.}~\bibnamefont{Erlenkaemper}},
  \bibnamefont{and} \bibinfo{author}{\bibfnamefont{K.}~\bibnamefont{Kruse}},
  \bibinfo{journal}{Phys. Rev. Lett.} \textbf{\bibinfo{volume}{108}}, \bibinfo{pages}{258103}
  (\bibinfo{year}{2012}).

\bibitem[{\citenamefont{Melbinger et~al.}(2012)\citenamefont{Melbinger, Reese,
  and Frey}}]{Melbinger:2012hp}
\bibinfo{author}{\bibfnamefont{A.}~\bibnamefont{Melbinger}},
  \bibinfo{author}{\bibfnamefont{L.}~\bibnamefont{Reese}}, \bibnamefont{and}
  \bibinfo{author}{\bibfnamefont{E.}~\bibnamefont{Frey}},
  \bibinfo{journal}{Phys. Rev. Lett.} \textbf{\bibinfo{volume}{108}}, \bibinfo{pages}{258104}
  (\bibinfo{year}{2012}).

\bibitem[{\citenamefont{Erlenk{\"a}mper and Kruse}(2013)}]{Erlenkamper:2013cy}
\bibinfo{author}{\bibfnamefont{C.}~\bibnamefont{Erlenk{\"a}mper}}
  \bibnamefont{and} \bibinfo{author}{\bibfnamefont{K.}~\bibnamefont{Kruse}},
  \bibinfo{journal}{J.~Chem.~Phys.}
  \textbf{\bibinfo{volume}{139}}, \bibinfo{pages}{164907}
  (\bibinfo{year}{2013}).

\bibitem[{\citenamefont{Mohapatra et~al.}(2015)\citenamefont{Mohapatra, Goode,
  and Kondev}}]{Mohapatra:2015cc}
\bibinfo{author}{\bibfnamefont{L.}~\bibnamefont{Mohapatra}},
  \bibinfo{author}{\bibfnamefont{B.~L.} \bibnamefont{Goode}}, \bibnamefont{and}
  \bibinfo{author}{\bibfnamefont{J.}~\bibnamefont{Kondev}},
  \bibinfo{journal}{PLoS Comput.~Biol.} \textbf{\bibinfo{volume}{11}}, \bibinfo{pages}{e1004160}
  (\bibinfo{year}{2015}).

\bibitem[{\citenamefont{Mohapatra et~al.}(2016)\citenamefont{Mohapatra, Goode,
  Jelenkovic, Phillips, and Kondev}}]{Mohapatra:2016kq}
\bibinfo{author}{\bibfnamefont{L.}~\bibnamefont{Mohapatra}},
  \bibinfo{author}{\bibfnamefont{B.~L.} \bibnamefont{Goode}},
  \bibinfo{author}{\bibfnamefont{P.}~\bibnamefont{Jelenkovic}},
  \bibinfo{author}{\bibfnamefont{R.}~\bibnamefont{Phillips}}, \bibnamefont{and}
  \bibinfo{author}{\bibfnamefont{J.}~\bibnamefont{Kondev}},
  \bibinfo{journal}{Annu.~Rev.~Biophys.} \textbf{\bibinfo{volume}{45}},
  \bibinfo{pages}{85} (\bibinfo{year}{2016}).

\bibitem[{\citenamefont{Fritzsche et~al.}(2016)\citenamefont{Fritzsche,
  Erlenk{\"a}mper, Moeendarbary, Charras, and Kruse}}]{Fritzsche:2016fx}
\bibinfo{author}{\bibfnamefont{M.}~\bibnamefont{Fritzsche}},
  \bibinfo{author}{\bibfnamefont{C.}~\bibnamefont{Erlenk{\"a}mper}},
  \bibinfo{author}{\bibfnamefont{E.}~\bibnamefont{Moeendarbary}},
  \bibinfo{author}{\bibfnamefont{G.}~\bibnamefont{Charras}}, \bibnamefont{and}
  \bibinfo{author}{\bibfnamefont{K.}~\bibnamefont{Kruse}},
  \bibinfo{journal}{Sci.~Adv.} \textbf{\bibinfo{volume}{2}},
  \bibinfo{pages}{e1501337} (\bibinfo{year}{2016}).

\bibitem[{\citenamefont{Channels et~al.}(2008)\citenamefont{Channels, Nedelec,
  Zheng, and Iglesias}}]{Channels:2008io}
\bibinfo{author}{\bibfnamefont{W.~E.} \bibnamefont{Channels}},
  \bibinfo{author}{\bibfnamefont{F.~J.} \bibnamefont{Nedelec}},
  \bibinfo{author}{\bibfnamefont{Y.}~\bibnamefont{Zheng}}, \bibnamefont{and}
  \bibinfo{author}{\bibfnamefont{P.~A.} \bibnamefont{Iglesias}},
  \bibinfo{journal}{Biophys.~J.} \textbf{\bibinfo{volume}{94}},
  \bibinfo{pages}{2598} (\bibinfo{year}{2008}).

\bibitem[{\citenamefont{Johann et~al.}(2015)\citenamefont{Johann, Goswami, and
  Kruse}}]{Johann:2015da}
\bibinfo{author}{\bibfnamefont{D.}~\bibnamefont{Johann}},
  \bibinfo{author}{\bibfnamefont{D.}~\bibnamefont{Goswami}}, \bibnamefont{and}
  \bibinfo{author}{\bibfnamefont{K.}~\bibnamefont{Kruse}},
  \bibinfo{journal}{Phys.~Rev.~Lett.} \textbf{\bibinfo{volume}{115}},
  \bibinfo{pages}{118103} (\bibinfo{year}{2015}).

\bibitem[{\citenamefont{Braun et~al.}(2017)\citenamefont{Braun, Lansky, Szuba,
  Schwarz, Mitra, Gao, L{\"u}decke, ten Wolde, and Diez}}]{Braun:2017hx}
\bibinfo{author}{\bibfnamefont{M.}~\bibnamefont{Braun}},
  \bibinfo{author}{\bibfnamefont{Z.}~\bibnamefont{Lansky}},
  \bibinfo{author}{\bibfnamefont{A.}~\bibnamefont{Szuba}},
  \bibinfo{author}{\bibfnamefont{F.~W.} \bibnamefont{Schwarz}},
  \bibinfo{author}{\bibfnamefont{A.}~\bibnamefont{Mitra}},
  \bibinfo{author}{\bibfnamefont{M.}~\bibnamefont{Gao}},
  \bibinfo{author}{\bibfnamefont{A.}~\bibnamefont{L{\"u}decke}},
  \bibinfo{author}{\bibfnamefont{P.~R.} \bibnamefont{ten Wolde}},
  \bibnamefont{and} \bibinfo{author}{\bibfnamefont{S.}~\bibnamefont{Diez}},
  \bibinfo{journal}{Nat.~Chem.~Biol.} \textbf{\bibinfo{volume}{268}},
  \bibinfo{pages}{9005} (\bibinfo{year}{2017}).

\bibitem[{\citenamefont{Prost et~al.}(2007)\citenamefont{Prost, Barbetta, and
  Joanny}}]{Prost2007}
\bibinfo{author}{\bibfnamefont{J.}~\bibnamefont{Prost}},
  \bibinfo{author}{\bibfnamefont{C.}~\bibnamefont{Barbetta}}, \bibnamefont{and}
  \bibinfo{author}{\bibfnamefont{J.-F.} \bibnamefont{Joanny}},
  \bibinfo{journal}{Biophys.~J.} \textbf{\bibinfo{volume}{93}},
  \bibinfo{pages}{1124} (\bibinfo{year}{2007}).

\bibitem[{\citenamefont{Loose et~al.}(2011)\citenamefont{Loose, Kruse, and
  Schwille}}]{Loose:2011dd}
\bibinfo{author}{\bibfnamefont{M.}~\bibnamefont{Loose}},
  \bibinfo{author}{\bibfnamefont{K.}~\bibnamefont{Kruse}}, \bibnamefont{and}
  \bibinfo{author}{\bibfnamefont{P.}~\bibnamefont{Schwille}},
  \bibinfo{journal}{Annu.~Rev.~Biophys.} \textbf{\bibinfo{volume}{40}},
  \bibinfo{pages}{315} (\bibinfo{year}{2011}).

\bibitem[{\citenamefont{Wettmann and Kruse}(2018)}]{Wettmann:2018bp}
\bibinfo{author}{\bibfnamefont{L.}~\bibnamefont{Wettmann}} \bibnamefont{and}
  \bibinfo{author}{\bibfnamefont{K.}~\bibnamefont{Kruse}},
  \bibinfo{journal}{Philos.~T.~Roy.~Soc.~B} \textbf{\bibinfo{volume}{373}},  \bibinfo{pages}{20170111}
  (\bibinfo{year}{2018}).

\bibitem[{\citenamefont{Laughlin}(2015)}]{Laughlin:2015hm}
\bibinfo{author}{\bibfnamefont{R.~B.} \bibnamefont{Laughlin}},
  \bibinfo{journal}{Proc.~Natl.~Acad.~Sci.} \textbf{\bibinfo{volume}{112}},
  \bibinfo{pages}{10371} (\bibinfo{year}{2015}).

\bibitem[{\citenamefont{Athamneh and Suter}(2015)}]{Athamneh:2015ba}
\bibinfo{author}{\bibfnamefont{A.~I.~M.} \bibnamefont{Athamneh}}
  \bibnamefont{and} \bibinfo{author}{\bibfnamefont{D.~M.} \bibnamefont{Suter}},
  \bibinfo{journal}{Front. Cell. Neurosci.} \textbf{\bibinfo{volume}{9}},
  \bibinfo{pages}{4481} (\bibinfo{year}{2015}).

\bibitem[{\citenamefont{Koser et~al.}(2016)\citenamefont{Koser, Thompson,
  Foster, Dwivedy, Pillai, Sheridan, Svoboda, Viana, Costa, Guck
  et~al.}}]{Koser:2016ks}
\bibinfo{author}{\bibfnamefont{D.~E.} \bibnamefont{Koser}},
  \bibinfo{author}{\bibfnamefont{A.~J.} \bibnamefont{Thompson}},
  \bibinfo{author}{\bibfnamefont{S.~K.} \bibnamefont{Foster}},
  \bibinfo{author}{\bibfnamefont{A.}~\bibnamefont{Dwivedy}},
  \bibinfo{author}{\bibfnamefont{E.~K.} \bibnamefont{Pillai}},
  \bibinfo{author}{\bibfnamefont{G.~K.} \bibnamefont{Sheridan}},
  \bibinfo{author}{\bibfnamefont{H.}~\bibnamefont{Svoboda}},
  \bibinfo{author}{\bibfnamefont{M.}~\bibnamefont{Viana}},
  \bibinfo{author}{\bibfnamefont{L.~d.~F.} \bibnamefont{Costa}},
  \bibinfo{author}{\bibfnamefont{J.}~\bibnamefont{Guck}}, \bibnamefont{et~al.},
  \bibinfo{journal}{Nat.~Neurosci.} \textbf{\bibinfo{volume}{19}},
  \bibinfo{pages}{1592} (\bibinfo{year}{2016}).

\bibitem[{\citenamefont{Erskine and Herrera}(2007)}]{Erskine:2007dw}
\bibinfo{author}{\bibfnamefont{L.}~\bibnamefont{Erskine}} \bibnamefont{and}
  \bibinfo{author}{\bibfnamefont{E.}~\bibnamefont{Herrera}},
  \bibinfo{journal}{Dev. Biol.} \textbf{\bibinfo{volume}{308}},
  \bibinfo{pages}{1} (\bibinfo{year}{2007}).

\bibitem[{\citenamefont{Albus et~al.}(2013)\citenamefont{Albus, Rishal, and
  Fainzilber}}]{Albus:2013es}
\bibinfo{author}{\bibfnamefont{C.~A.} \bibnamefont{Albus}},
  \bibinfo{author}{\bibfnamefont{I.}~\bibnamefont{Rishal}}, \bibnamefont{and}
  \bibinfo{author}{\bibfnamefont{M.}~\bibnamefont{Fainzilber}},
  \bibinfo{journal}{Trends Cell Biol.} \textbf{\bibinfo{volume}{23}},
  \bibinfo{pages}{305} (\bibinfo{year}{2013}).

\bibitem[{\citenamefont{Rishal et~al.}(2012)\citenamefont{Rishal, Kam, Perry,
  Shinder, Fisher, Schiavo, and Fainzilber}}]{Rishal:2012jg}
\bibinfo{author}{\bibfnamefont{I.}~\bibnamefont{Rishal}},
  \bibinfo{author}{\bibfnamefont{N.}~\bibnamefont{Kam}},
  \bibinfo{author}{\bibfnamefont{R.~B.-T.} \bibnamefont{Perry}},
  \bibinfo{author}{\bibfnamefont{V.}~\bibnamefont{Shinder}},
  \bibinfo{author}{\bibfnamefont{E.~M.~C.} \bibnamefont{Fisher}},
  \bibinfo{author}{\bibfnamefont{G.}~\bibnamefont{Schiavo}}, \bibnamefont{and}
  \bibinfo{author}{\bibfnamefont{M.}~\bibnamefont{Fainzilber}},
  \bibinfo{journal}{Cell Rep.} \textbf{\bibinfo{volume}{1}},
  \bibinfo{pages}{608} (\bibinfo{year}{2012}).

\bibitem[{\citenamefont{Perry et~al.}(2016)\citenamefont{Perry, Rishal,
  Doron-Mandel, Kalinski, Medzihradszky, Terenzio, Alber, Koley, Lin, Rozenbaum
  et~al.}}]{Perry:2016he}
\bibinfo{author}{\bibfnamefont{R.~B.-T.} \bibnamefont{Perry}},
  \bibinfo{author}{\bibfnamefont{I.}~\bibnamefont{Rishal}},
  \bibinfo{author}{\bibfnamefont{E.}~\bibnamefont{Doron-Mandel}},
  \bibinfo{author}{\bibfnamefont{A.~L.} \bibnamefont{Kalinski}},
  \bibinfo{author}{\bibfnamefont{K.~F.} \bibnamefont{Medzihradszky}},
  \bibinfo{author}{\bibfnamefont{M.}~\bibnamefont{Terenzio}},
  \bibinfo{author}{\bibfnamefont{S.}~\bibnamefont{Alber}},
  \bibinfo{author}{\bibfnamefont{S.}~\bibnamefont{Koley}},
  \bibinfo{author}{\bibfnamefont{A.}~\bibnamefont{Lin}},
  \bibinfo{author}{\bibfnamefont{M.}~\bibnamefont{Rozenbaum}},
  \bibnamefont{et~al.}, \bibinfo{journal}{Cell Rep.}
  \textbf{\bibinfo{volume}{16}}, \bibinfo{pages}{1664} (\bibinfo{year}{2016}).

\bibitem[{\citenamefont{Myers and Baas}(2007)}]{Myers:2007ig}
\bibinfo{author}{\bibfnamefont{K.~A.} \bibnamefont{Myers}} \bibnamefont{and}
  \bibinfo{author}{\bibfnamefont{P.~W.} \bibnamefont{Baas}},
  \bibinfo{journal}{J. Cell Biol.} \textbf{\bibinfo{volume}{178}},
  \bibinfo{pages}{1081} (\bibinfo{year}{2007}).

\bibitem[{\citenamefont{Karamched and Bressloff}(2015)}]{Karamched:2015ha}
\bibinfo{author}{\bibfnamefont{B.~R.} \bibnamefont{Karamched}}
  \bibnamefont{and} \bibinfo{author}{\bibfnamefont{P.~C.}
  \bibnamefont{Bressloff}}, \bibinfo{journal}{Biophys.~J.}
  \textbf{\bibinfo{volume}{108}}, \bibinfo{pages}{2408} (\bibinfo{year}{2015}).

\bibitem[{\citenamefont{Bressloff and Karamched}(2015)}]{Bressloff:2015gg}
\bibinfo{author}{\bibfnamefont{P.~C.} \bibnamefont{Bressloff}}
  \bibnamefont{and} \bibinfo{author}{\bibfnamefont{B.~R.}
  \bibnamefont{Karamched}}, \bibinfo{journal}{Front. Cell. Neurosci.}
  \textbf{\bibinfo{volume}{9}} (\bibinfo{year}{2015}).

\bibitem[{\citenamefont{Bray}(1979)}]{Bray:1979tg}
\bibinfo{author}{\bibfnamefont{D.}~\bibnamefont{Bray}}, \bibinfo{journal}{J.~Cell Sci.} 
\textbf{\bibinfo{volume}{37}}, \bibinfo{pages}{391}
  (\bibinfo{year}{1979}).

\bibitem[{\citenamefont{Ahmad et~al.}(2000)\citenamefont{Ahmad, Hughey,
  Wittmann, Hyman, Greaser, and Baas}}]{Ahmad:2000wv}
\bibinfo{author}{\bibfnamefont{F.~J.} \bibnamefont{Ahmad}},
  \bibinfo{author}{\bibfnamefont{J.}~\bibnamefont{Hughey}},
  \bibinfo{author}{\bibfnamefont{T.}~\bibnamefont{Wittmann}},
  \bibinfo{author}{\bibfnamefont{A.}~\bibnamefont{Hyman}},
  \bibinfo{author}{\bibfnamefont{M.}~\bibnamefont{Greaser}}, \bibnamefont{and}
  \bibinfo{author}{\bibfnamefont{P.~W.} \bibnamefont{Baas}},
  \bibinfo{journal}{Nat.~Cell Biol.} \textbf{\bibinfo{volume}{2}},
  \bibinfo{pages}{276} (\bibinfo{year}{2000}).

\bibitem[{\citenamefont{Bernal et~al.}(2007)\citenamefont{Bernal, Pullarkat,
  and Melo}}]{Bernal:2007gx}
\bibinfo{author}{\bibfnamefont{R.}~\bibnamefont{Bernal}},
  \bibinfo{author}{\bibfnamefont{P.~A.} \bibnamefont{Pullarkat}},
  \bibnamefont{and} \bibinfo{author}{\bibfnamefont{F.}~\bibnamefont{Melo}},
  \bibinfo{journal}{Phys. Rev. Lett.} \textbf{\bibinfo{volume}{99}},
  \bibinfo{pages}{018301} (\bibinfo{year}{2007}).

\bibitem[{\citenamefont{Baas and Ahmad}(2001)}]{Baas:2001kk}
\bibinfo{author}{\bibfnamefont{P.~W.} \bibnamefont{Baas}} \bibnamefont{and}
  \bibinfo{author}{\bibfnamefont{F.~J.} \bibnamefont{Ahmad}},
  \bibinfo{journal}{Trends Cell Biol.} \textbf{\bibinfo{volume}{11}},
  \bibinfo{pages}{244} (\bibinfo{year}{2001}).

\bibitem[{\citenamefont{Alon}(2006)}]{Alon06:nn}
\bibinfo{author}{\bibfnamefont{U.}~\bibnamefont{Alon}},
  \emph{\bibinfo{title}{{An introduction to systems biology: design principles
  of biological circuits}}} (\bibinfo{year}{2006}).

\bibitem[{\citenamefont{Mueller et~al.}(2017)\citenamefont{Mueller, Szep,
  Nemethova, de~Vries, Lieber, Winkler, Kruse, Small, Schmeiser, Keren
  et~al.}}]{Mueller:2017br}
\bibinfo{author}{\bibfnamefont{J.}~\bibnamefont{Mueller}},
  \bibinfo{author}{\bibfnamefont{G.}~\bibnamefont{Szep}},
  \bibinfo{author}{\bibfnamefont{M.}~\bibnamefont{Nemethova}},
  \bibinfo{author}{\bibfnamefont{I.}~\bibnamefont{de~Vries}},
  \bibinfo{author}{\bibfnamefont{A.~D.} \bibnamefont{Lieber}},
  \bibinfo{author}{\bibfnamefont{C.}~\bibnamefont{Winkler}},
  \bibinfo{author}{\bibfnamefont{K.}~\bibnamefont{Kruse}},
  \bibinfo{author}{\bibfnamefont{J.~V.} \bibnamefont{Small}},
  \bibinfo{author}{\bibfnamefont{C.}~\bibnamefont{Schmeiser}},
  \bibinfo{author}{\bibfnamefont{K.}~\bibnamefont{Keren}},
  \bibnamefont{et~al.}, \bibinfo{journal}{Cell} \textbf{\bibinfo{volume}{171}},
  \bibinfo{pages}{188} (\bibinfo{year}{2017}).

\bibitem[{\citenamefont{Kruse et~al.}(2004)\citenamefont{Kruse, Joanny,
  J{\"u}licher, Prost, and Sekimoto}}]{Kruse:2004il}
\bibinfo{author}{\bibfnamefont{K.}~\bibnamefont{Kruse}},
  \bibinfo{author}{\bibfnamefont{J.-F.} \bibnamefont{Joanny}},
  \bibinfo{author}{\bibfnamefont{F.}~\bibnamefont{J{\"u}licher}},
  \bibinfo{author}{\bibfnamefont{J.}~\bibnamefont{Prost}}, \bibnamefont{and}
  \bibinfo{author}{\bibfnamefont{K.}~\bibnamefont{Sekimoto}},
  \bibinfo{journal}{Phys. Rev. Lett.} \textbf{\bibinfo{volume}{92}},
  \bibinfo{pages}{078101} (\bibinfo{year}{2004}).

\bibitem[{\citenamefont{Bois et~al.}(2011)\citenamefont{Bois, J{\"u}licher, and
  Grill}}]{Bois:2011kx}
\bibinfo{author}{\bibfnamefont{J.~S.} \bibnamefont{Bois}},
  \bibinfo{author}{\bibfnamefont{F.}~\bibnamefont{J{\"u}licher}},
  \bibnamefont{and} \bibinfo{author}{\bibfnamefont{S.~W.} \bibnamefont{Grill}},
  \bibinfo{journal}{Phys.~Rev.~Lett.} \textbf{\bibinfo{volume}{106}},
  \bibinfo{pages}{028103} (\bibinfo{year}{2011}).

\bibitem[{Note1()}]{Note1}
\bibinfo{note}{These results also hold if the value of $J_I$ is fixed
  while changing $J_0$.}


\bibitem[{\citenamefont{Crook et~al.}(2003)\citenamefont{Crook, Scheper, and
  Pathirana}}]{Crook:2003ci}
\bibinfo{author}{\bibfnamefont{N.}~\bibnamefont{Crook}},
  \bibinfo{author}{\bibfnamefont{T.~O.} \bibnamefont{Scheper}},
  \bibnamefont{and}
  \bibinfo{author}{\bibfnamefont{V.}~\bibnamefont{Pathirana}},
  \bibinfo{journal}{Information Sciences} \textbf{\bibinfo{volume}{150}},
  \bibinfo{pages}{59} (\bibinfo{year}{2003}).

\end{thebibliography}
\end{document}